\documentclass[10pt,aps,pre,twocolumn,superscriptaddress]{revtex4-2}
\usepackage[linkcolor = blue, citecolor = red, urlcolor = blue, colorlinks = true]{hyperref}
\usepackage[usenames,dvipsnames,x11names]{xcolor}
\usepackage[normalem]{ulem}
\usepackage{graphicx}
\usepackage{listings}
\usepackage{stmaryrd}
\usepackage{amssymb}
\usepackage{amsmath}
\usepackage{gensymb}
\usepackage{comment}
\usepackage{float}
\usepackage{bbold}
\usepackage{bm}
\definecolor{crimson}{rgb}{0.75686,0,0.262745}
\definecolor{saphire}{rgb}{0.0,0.196,0.372549}
\definecolor{plum}{rgb}{0.50588,0.007843,0.3843137}

\begin{document}

\title{Thinning-by-spinning: shear rheology of dense chiral fluids}

    \author{Lucio Mauro Carenza}
    
    \affiliation{Dipartimento  Interateneo di  Fisica,        Universit\`a  degli  Studi  di  Bari  and  INFN,
        Sezione  di  Bari,  via  Amendola  173,  Bari,  I-70126,  Italy}

\author{Giuseppe Gonnella}
    \affiliation{Dipartimento  Interateneo di  Fisica,        Universit\`a  degli  Studi  di  Bari  and  INFN,
        Sezione  di  Bari,  via  Amendola  173,  Bari,  I-70126,  Italy}
\author{Demian Levis}
\email{levis@ub.edu}

\affiliation{Computing and Understanding Collective Action (CUCA) Lab, Condensed Matter Physics Department, Facultat de Física, Universitat de Barcelona, Martí i Franquès 1,
E08028 Barcelona, Spain}

    \affiliation{UBICS, University of Barcelona Institute     of Complex Systems, Mart\'i i Franqu\`es 1, E08028   Barcelona, Spain}
\author{Giuseppe Negro}
\email{giuseppe.negro@ed.ac.uk}
\affiliation{SUPA, School of Physics and Astronomy,      University of Edinburgh, Edinburgh EH9 3FD, UK}

\begin{abstract}
We investigate the linear and nonlinear rheology of dense chiral fluids composed of self-spinning particles under external shear. Using particle-based simulations of a two-dimensional Lennard-Jones model with transverse interactions, we show that chirality acts as an intrinsic source of fluctuations and shear. In the solid regime, spinning fluidizes the system, weakening hexatic order. In the liquid regime, the viscosity is quantitatively described by a Green-Kubo relation upon replacing the temperature by a chirality-dependent effective temperature. Beyond linear response, flow curves collapse when expressed in terms of the ratio between imposed shear and spinning rates, revealing a thinning-by-spinning mechanism. At large forcing, this correspondence breaks down and a pronounced handedness asymmetry emerges: when transverse interactions oppose the imposed shear, stresses relax through the formation of string-like flow channels. Our results identify chirality as a generic mechanism for fluidization and provide a unified framework for understanding the rheology of dense chiral suspensions.
\end{abstract}

\maketitle

Chiral fluids composed of self-spinning particles provide a versatile platform for realizing complex non-equilibrium states of matter. In such systems, spinning generically generates odd, or transverse, forces that reshape the phase behavior and dynamics of soft matter \cite{ 
han2021fluctuating, Liebchen2022, fruchart2023odd}. Transverse forces can arise from hydrodynamic flows induced by rotating objects at low Reynolds number, as in the case of  starfish embryos at fluid interfaces ~\cite{smith2019,
tan2022livingcrystals}, bacteria  \cite{petroff2015fast, li2024robust}, sperm cells \cite{riedel2005self}, or monolayers of magnetic colloids driven by a rotating magnetic field \cite{yan2015rotatingcrystals,coughlan2016rotatingcolloids,wang2019spinners, soni2019odd,grzybowski2000dynamic, massana2021arrested}. At the granular scale, transverse forces typically arise from frictional contacts ~\cite{scholz2018rotatingrobots, 
yang2020robust, arora2021emergent, caprini2025active, carrillo2025depinning}, and, quite generally,  interactions between non-spherical spinning objects have a transverse component, translating the breakdown of parity symmetry \cite{nguyen2014emergent, Zuiden2016, hargus2020time,  poggioli2023odd}. In this context, a fundamental question that remains largely open is how the properties of the different phases of matter, such as the fluid and solid, exhibited by assemblies of particles, are affected by self-spinning.

\begin{figure}[h!]
    \centering
\includegraphics[width=0.9\columnwidth]{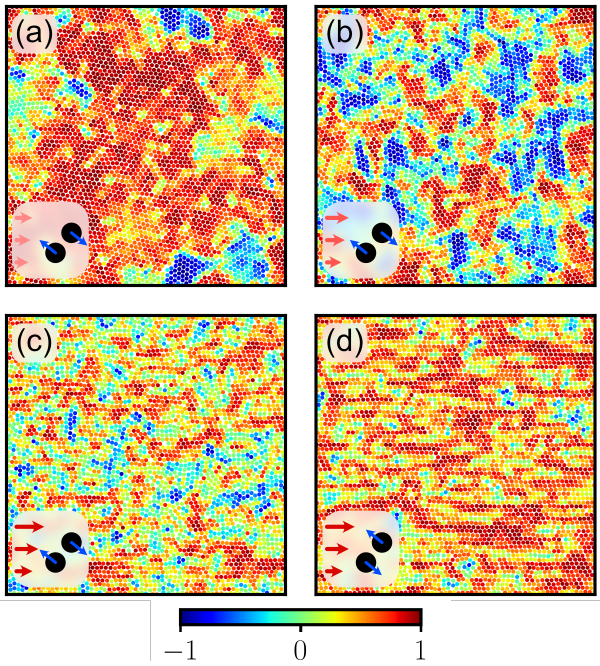}
    \caption{Projection of the local hexatic order parameter $\psi_6(r_i)$ on the global mean orientation at fixed  density $\rho=0.7$ and temperature $k_BT=0.35$ for different values of shear rate and chirality.
    Panels correspond to   $\Omega=5$ (a)  $\dot\gamma=10^{-4}$, (b)   $\dot\gamma = 10^{-2}$ and (c) $\dot\gamma=10^{-1}$.  Panel (d) corresponds to  $\Omega=-5$ and $\dot\gamma=10^{-1}$.
    }
    \label{fig1}
\end{figure}

In the absence of transverse forces, two-dimensional ($2d$)  dense particle systems   typically exhibit hexatic and solid phases~\cite{ryzhov2017berezinskii, Dijkstra2020,Digregorio2022}, which can be fluidized by shear or, in thermal systems, by increasing temperature. The associated viscosity is in general a non-linear function of the shear rate, growing with density and eventually diverging when solidity emerges \cite{bonn2017yield}.  At low shear rates, a fluid is in a linear reponse  regime, where the viscosity is provided by a Green-Kubo relation. It is typically followed by a non-linear,  shear-thinning regime, at high shear rates, characterized by a decrease of the viscosity with the imposed drive \cite{mewis2012colloidal}. 

A growing body of experiments and simulations has shown that spinning can also fluidize an otherwise  $2d$  solid, breaking it  into a dense  fluid composed of transient hexatic rotating domains separated by evolving grain boundaries  ~\cite{Zuiden2016, Koko2017, Suchla2021, bililign2022motile, caporusso2024}.  Like  shear, self-spinning generates mutual torques that drive local rearrangements and promote defect proliferation, eroding positional and bond-orientational order (see Fig. \ref{fig1}).
This raises two natural questions: is self-spinning equivalent to a local shear, and how does a chiral fluid respond to an externally imposed shear? 

To address this problem, we investigate the rheology of a model chiral fluid, build as a chiral extension of the Lennard-Jones model in $2d$ ~\cite{barker1981} with pairwise  transverse forces~\cite{caporusso2024}.   This system isolates the roles of transverse forces and global shear while remaining continuously connected to a well-known equilibrium reference system.
We establish its entire flow curves, both as a function of applied shear and the spinning rate. 
We find that the chiral system behaves as a sheared fluid with a chirality-dependent effective temperature. By increasing the transverse forces,  the shear viscosity of the system decreases, a phenomenon that we coin thinning-by-spinning, while the effective temperature (extracted from the violations of the fluctuation-dissipation theorem) increases. 
Chirality and external driving reinforce each other, producing shear thinning behaviour across both the solid and fluid phases. 
Moreover, as shown in Fig. \ref{fig1}d, when the handedness of the imposed shear is opposite to the particles' one, at high enough shear rates the chiral fluid self-organizes into stripes, exhibiting enhanced  positional  order as observed in dense systems at larger shear rates  \cite{Ghimenti2024}. Here the onset and character of such stripes is controlled by the strength and handedness of the transverse forces.

    \textit{Model-} We consider  a system of $N=4096$ spherical particles in $2d$ (disks) located at $\{\bm{r}_i=(x_i,y_i)\}_{i=1}^N$ in a a rectangular box of area  $A=L_x L_y$  such that $L_y/L_x = \sin(\pi/3)$ 
 with Lees-Edwards boundary conditions \cite{allen2017computer}. The dynamics of  particle $i$ obeys
    \begin{equation}\label{eq:eom}
        m \ddot {\bm r}_i + \Gamma\dot{\bm r}_i = \sum_{j\neq i} \left[\bm{f}_{ij} - \bm{\nabla}_i U(r_{ij})\right] + \Gamma \dot\gamma y_i \bm{\hat{x}} + \sqrt{2\Gamma k_B T} \bm{\nu}_i,
    \end{equation} 
where $\Gamma$ is the damping coefficient, $\dot\gamma$ the shear rate, and $\bm{\nu}_i$ a Gaussian white noise with $\left<\nu_{i\alpha}(t) \nu_{j\beta}(t')\right> = \delta_{ij}\delta_{\alpha\beta}\delta(t-t')$ and zero mean.  Conservative interactions derive from the Lennard-Jones potential $U$,  with typical length and energy scale  $\sigma_d$ and $\epsilon$. Non-conservative transverse forces are introduced as $\bm{f}_{ij} = -\Gamma \bm{u}_{ij} $,  where $\bm {u}_{ij} = \omega \sigma_d^3\hat{z}\times\frac{\bm{r}_{ij}}{r_{ij}^3}$, and $\bm{r}_{ij}=\bm{r}_i-\bm{r}_j$. Both $U$ and $\bm{f}_{ij}$ are truncated at $r_c=5\sigma_d$.  We fix $m=\sigma_d=\epsilon=1$ and $\Gamma=100$, fixing the  unit of time  $\tau=\sqrt{m\sigma_d^2/\epsilon}$.  We explore shear rates $10^{-5}\leq \dot\gamma \leq 1$. The non-dimensional parameter $\Omega=\Gamma\omega\tau^2/m$.  
For $\Omega>0$ the transverse forces' torque are aligned with the vorticity of the imposed shear,  while for $\Omega<0$ they are opposite. We then integrate eq. \ref{eq:eom} using a velocity-Verlet scheme \cite{thompson2022lammps}.

In the absence of shear, the model displays a  phase behavior featuring  a coexistence region between a dilute gas (G) and a chiral liquid (CL), followed by a (homogeneous) chiral liquid (CL)
and finally a  solid phase (S). The latter is  characterized by a non-zero value of the global hexatic order parameter $\Psi_6=\sum_i^N\psi_6(r_i)/N$, 
 where $\psi_6(r_j)=n_j^{-1}\sum _{k\in\partial_j}e^{i6\theta_{jk}}$ is the local hexatic field (the sum is taken across the Voronoi neighbours, out of the $n_j$ ones of a particle '$j$',  $\theta_{jk}$ being the  bond angle formed by the direction connecting the center of the $j$th disk  to its $k$th nearest neighbour and $\bm{\hat{x}})$. 
    
    \textit{Thinning by spinning}- We consider the system in its chiral fluid phase. As depicted in Fig. \ref{fig1},  hexatic order is altered upon shearing. As $\dot\gamma$ increases  large hexatic regions break into  patches of local hexatic order,  decreasing in size as the system is further driven. This is reminiscent of the melting induced by chirality observed in the absence of external driving~\cite{caporusso2024,bililign2022motile}, thus suggesting an analogy between   shear and chiral forces. 
As shown in Fig. \ref{fig1}c-d, for $\Omega<0$, the system re-structures into string-like ordered structures, in a situation  where,   at $\Omega>0$ (but same value  $|\Omega|=5$),  it exhibits hexatic patches moderately elongated along the direction of shear. Since such quasi-1D structures emerge in dense passive systems for large driving \cite{Ghimenti2024}, one can think that the effect of chirality when its handedness is opposite to the shear one, is to further increase the overall effective shear.

To quantify the shear rheology of the model, 
we compute the Irving--Kirkwood stress tensor $\bm{\sigma}$, as given by:
    \begin{equation}
        \sigma_{\alpha\beta} = -\frac{1}{A} \left< \sum_i m \dot{r}_{i,\alpha}\dot{r}_{i,\beta} + \sum_{i<j} r_{ij,\alpha} F_{ij,\beta}\right>,
    \end{equation}
where latin indices label particles, greek ones cartesian components and $\bm{F}_{ij}=\bm{f}_{ij} - \bm{\nabla}_i U(r_{ij})$.  
Due to the transverse forces, $\sigma$ contains an antisymmetric component associated with the rotational viscosity \cite{caporusso2024}. We thus define the shear viscosity as 
\begin{equation}\label{eq:visc}
    \eta= \frac{\left< \sigma_{xy}+\sigma_{yx}\right>} {2 \dot\gamma} \equiv \frac{\left<\sigma^s\right>}{\dot\gamma} , 
\end{equation}
where $\sigma^{\rm s}$
is the symmetrized stress tensor, removing the purely rotational contribution.  
    
\begin{figure}[t!]
    \centering
    \includegraphics[width=1.\columnwidth]{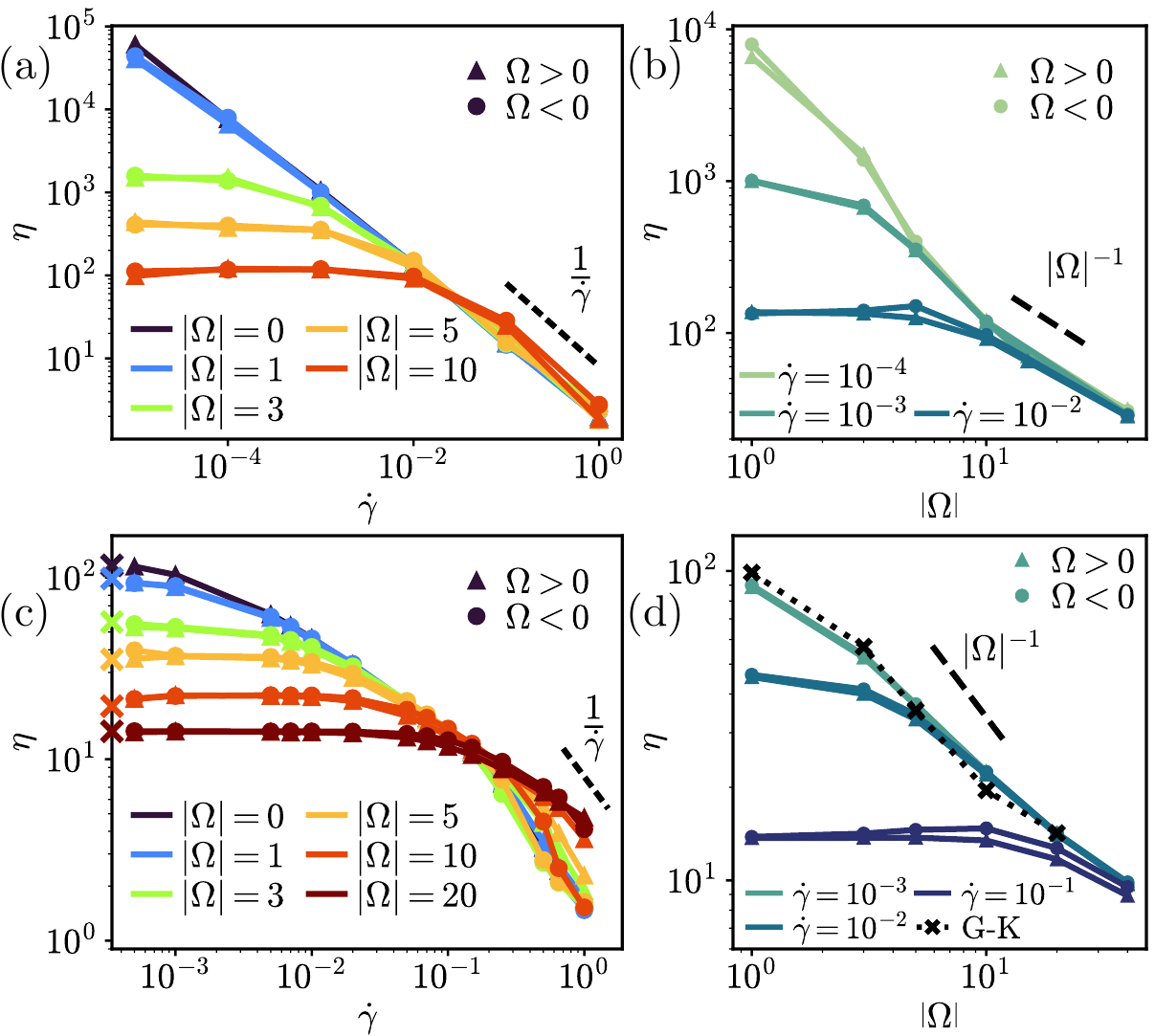}
    \caption{
    (a) Shear viscosity $\eta$ for $\rho=0.7$ and $k_BT=0.35$ (in the solid phase in equilibrium) as a function of the shear rate, $\dot\gamma$, for different values of  $|\Omega|$ (both positive and negative). (b) $\eta$ for the same system, now as a function of $|\Omega|$, highlighting the thinning-by-spinning. (c) 
    Viscosity for $\rho=0.6$ and $k_BT=0.47$ (in the fluid phase for the passive system), compared with values obtained from the Green-Kubo relation (crosses on the vertical axis). (d) Thinning by spinning and comparison with Green-Kubo relation (blck crosses). 
    Dashed lines show the scaling   $\sim1/\dot\gamma$ and $\sim1/|\Omega|$ as guides to the eye.}  
    \label{fig2}
\end{figure} 
    
    Starting from an equilibrium state deep within the solid phase ($\rho=0.7$ and $k_BT=0.35)$, Fig.\ref{fig2}a shows the flow curves for different  $|\Omega|$. The equilibrium system ($\Omega=0$) does not flow at low shear rates, signaled by the divergence of $\eta$. Beyond a threshold value of $\Omega$, the system is fluidized, akin a yield stress, that one identifies by the emergence of a finite $\eta$ in the zero shear limit.  As chirality is further increased, the Newtonian regime over which the shear stress grows  linearly with $\dot{\gamma}$, is enlarged, covering a broader range of values.  
Beyond this linear regime, we found that both chirality and shear reduce the viscosity,  driving the system into a shear thinning regime, with a  scaling compatible with the Mode-Coupling Theory prediction,   $\eta\sim 1/\dot \gamma$ \cite{fuchs2002theory},  for all considered values of $\Omega$ . 
As anticipated by the inspection of the snapshots Fig. \ref{fig1}, and shown quantitatively in Fig. \ref{fig2}c, when $\Omega<0$ the system responds as if the shear rate was larger: the viscosity gets smaller in the non-linear regime, while no significant differences are observed in the linear one. 

To disentangle the role played by chirality from the imposed shear flow,  we also  compute the viscosity at  fixed  $\dot{\gamma}$  but varying  $|\Omega|$.  As depicted in Fig.\ref{fig2}b-d, the viscosity monotonically decreases with  $|\Omega|$, highlighting a macroscopic thinning effect induced purely by the microscopic nature of the pairwise chiral interactions.   
    These observations go in the direction of interpreting chirality as a source of shear stresses, inducing a analog of shear thinning, with a scaling comparable to $\eta \sim \Omega^{-1}$, although purely induced by spinning.

\textit{Scaling and effective temperature-}
    So far we have seen that the viscosity decreases with both increasing shear rate and chirality. 
    We now ask whether these two dependencies can be rationalized in a unified way. 
    To this end, we compare the non-equilibrium shear  viscosity with predictions from equilibrium linear response theory.
\begin{figure}[t!]
    \centering
    \includegraphics[width=1.\columnwidth]{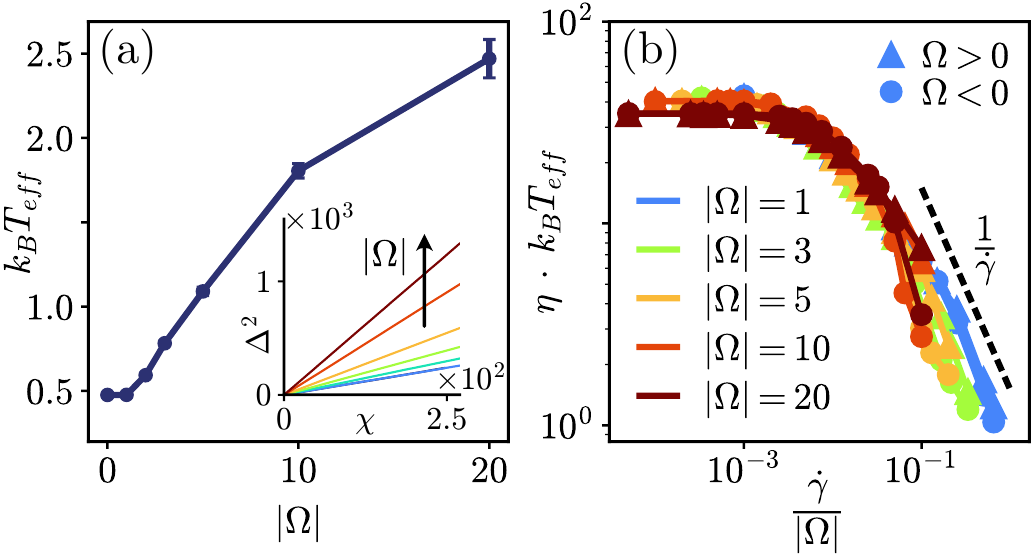}
    \caption{
    (a) Measure of the effective temperature of the system as a function of the activity.
    The inset shows the parametric plot $\Delta^2(\chi)$ for the different values of activity, $|\Omega|$, shown in the main plot. 
    (b) Flow curves   at $\rho=0.6$ and $k_BT=0.47$ after rescaling the viscosity by $T_{\rm eff}$ and the shear rate by $|\Omega|$. 
    } \label{fig3}
\end{figure}

    We compute an effective temperature $T_{\rm eff}$, defined via the violations of the Fluctuation-Dissipation Theorem (FDT), in particular, the Stokes-Einstein relation $k_BT_{\rm eff} = D/\mu$. 
    To do so, we apply a  constant force $\bm{f}_i = \epsilon_i f_0 \bm{e}_x$ to each particle, where $\epsilon_i = \pm 1$ with equal probability. We choose $f_0$ to be small enough to lie in the linear response regime (not to significantly affect the  dynamics of the system). 
    We then compute the integrated response function
    $\chi = \lim_{f_0 \to 0} N^{-1} \sum_{i=1}^N {\epsilon_i}\langle x_i(t) - x_i(0)\rangle /{f_0} $, from which one extracts the mobility  from the linear drift induced by the applied  perturbation as $\mu = \lim_{t\to \infty}\chi(t)/ t$ ~\cite{levis2015single, dal2019linear, petrelli2020effective}.
   The diffusion coefficient $D$  is extracted from the long-time behaviour of the mean-squared displacement $\Delta^2(t) = N^{-1} \sum_{i=1}^N \langle (x_i(t) - x_i(0))^2\rangle\to_{t\to\infty} 2Dt$,  in unperturbed conditions (see  Fig.~\ref{fig3} a).  
   As shown in Fig.~\ref{fig3}a, $T_{\rm eff}$ grows monotonically with $\Omega$, confirming that chiral activity injects energy in a manner that can be captured at large-scales by an effective temperature.
Equipped with this independently measured $T_{\rm eff}$, we propose the following generalization of the Green--Kubo expression to account for the linear response of the chiral system, as done for dense systems of self-propelled particles \cite{wiese2023fluid}:
\begin{equation}
    \eta = \frac{A}{k_BT_{\rm eff}(\Omega)} \int_0^{+\infty} 
    \langle \sigma^{\rm s}(0)\,\sigma^{\rm s}(t)
    \rangle_0\, dt,
    \label{eq:GK}
\end{equation}
    where stress correlations are taken over unperturbed dynamics and $A$ is the system area. The Green--Kubo integral, rescaled by  a chirality dependent  $T_{\rm eff}$, shows remarkable agreement with the macroscopic viscosity, independently computed from eq. \ref{eq:visc} in the $\dot{\gamma}\to 0$ limit (crosses in Fig.~\ref{fig2}c,d). 
    By multiplying the viscosity by $T_{\rm eff}(\Omega)$ and dividing the shear rate by $|\Omega|$,  all flow curves collapse to a single master curve (up to large values of $\dot{\gamma}$, where the $\Omega>0$ and $\Omega<0$ branches start deviating). 
Such scaling, shows that increasing chirality is like increasing the shear rate of an otherwise equilibrium dense fluid, at a higher effective temperature, over a broad parameter range. At larger values of the drive, the sign of $\Omega$ starts to matter and a new regime, characterized by a structural change, sets  in.

\begin{figure}[t!]
    \centering
    \includegraphics[width=1\columnwidth]{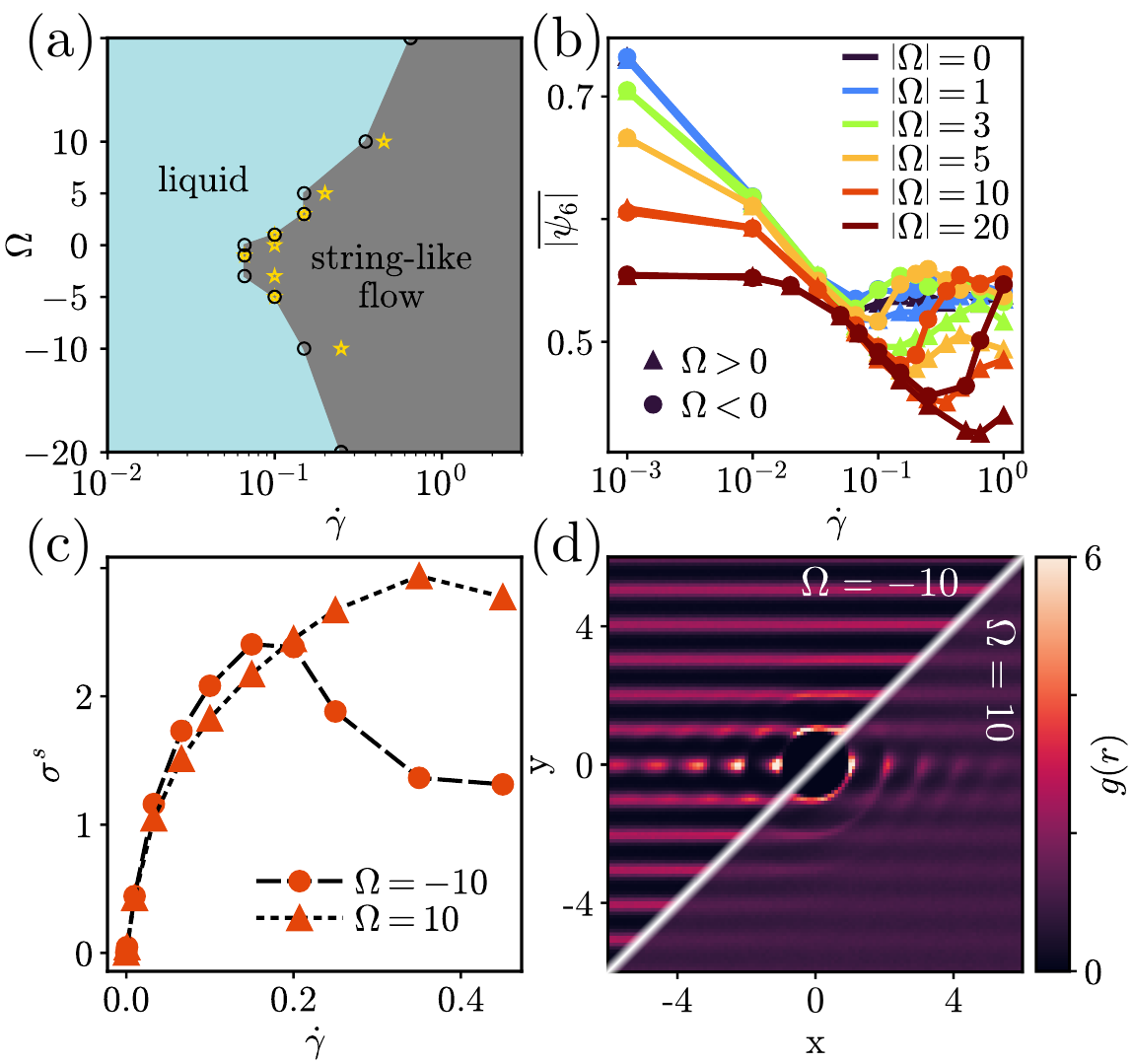}
    \caption{\textbf{Phase diagram and layering.} (a) Phase diagram of the transition to layering for $\rho=0.65$ and $k_BT=0.35$. The liquid-layering transition is identified by the minimum in   $\overline{|\psi_6|}$, 
    $\dot\gamma^*(\Omega)$  (empty circles) and by the onset of an order-of-magnitude increase in the 
    peak of the structure factor (gold stars). (b) Mean value of the hexatic order parameter for the same density and temperature. (c) Measured off diagonal stress for $\Omega=10$ and $\Omega=-10$ as a function of $\dot\gamma$. (d) Pair correlation functions at $\dot\gamma=0.35$, for $\Omega=10$ (liquid regime) and $\Omega=-10$ (string-like regime),  
    highlighting the asymmetry in $\Omega$ in the transition from a liquid to a string-like flow. 
    }
    
\label{fig4}
\end{figure}

\textit{Chirality induced layering-}
    In the strongly driven regime the map of local hexatic order parameter, Fig.~\ref{fig1}c-d,  reveals a structural difference between the two handednesses.
    When chirality  is opposed to the imposed shear, the system develops elongated string-like ordered domains along which particles'  flow is facilitated, as observed in highly driven  systems of passive particles~\cite{Ghimenti2024}.

    A structural analysis in the  $(\dot\gamma, \Omega)$ plane allows to construct the state diagram  Fig.~\ref{fig4}a. At low shear rates (liquid regime), the two handednesses are nearly indistinguishable at the macroscopic level. Here chirality acts as an extra source of shear, accompanied by an increasing  $T_{\rm eff}$.  
At larger drives the system gets fluidized by the emergence of quasi-1D structures, that are promoted for $\Omega<0$. 
To quantitatively identify such different flow regimes, we rely on structural quantities, such as the average of the module of local hexatic order, via $\overline{|\psi_6|} =\sum_i |\psi_6(r_i)|/N $,  the pair correlation function $g(x,y)$, showed in Fig.\ref{fig4}b,d,  and the structure factor (see End Matter).
At small shear rates, the hexatic order not  distinguishes  $\Omega>0$ and $\Omega<0$.  Reversing the sense of rotation does not produce an observable structural difference.

As $\dot\gamma$ is further increased, $\overline{|\psi_6|}$ exhibits a non-monotonic behaviour, developing a  minimum  at a given value $\dot\gamma=\dot\gamma^*$ that can be identified as the  onset of the string-like flow regime. Such a layered structure of  well packed particles along strings,  translates into  an increase in the overall hexatic order.  The  value $\dot\gamma^*$ grows as $|\Omega|$ increases, and it is smaller for negative values. 
Starting from a system in the string-like flow regime, chirality renders the flow less localized along quasi-1D structures, to eventually make it  liquid-like at larger values of $\Omega$.    

The impact of the sign of $\Omega$ is further reflected in the  mechanical response. As shown in  Fig.~\ref{fig4}c, the off-diagonal stresses measured for $\Omega = 10$ and $\Omega = -10$ diverge significantly as $\dot\gamma>\dot\gamma^*$: the two branches develop opposite trends beyond the onset of string-like flow. For $\Omega<0$, the stresses get reduced beyond  $\dot\gamma^*\approx 0.2$ by the presence of strings,  along which particles flow more easily. 
On the contrary, for $\Omega>0$ one has to go up to $\dot\gamma^*\approx 0.4$ to eventually observe the formation of such strings and the resulting relaxation of stresses.  

To further characterize such structural difference between systems with opposite handedness, we display in Fig.~\ref{fig4}d, the pair correlation function. 
For the system at $\Omega=-10$ it shows a clear signal of layering, with strong peaks along $\bm{\hat{x}}$, and a periodic signal along $\bm{\hat{y}}$. Such structural signatures are washed out at $\Omega=10$ for the same shear rate. The competition between the imposed chirality of the shear and the intrinsic one of the particles, induces and enhanced shear thinning and layering.

\textit{Discussion and conclusions--}
We have presented a comprehensive characterization of the shear rheology of a paradigmatic model of chiral particles in which chirality is encoded through pairwise transverse interactions, mimicking the mechanical effect of spinning dissipative objects.  
In the absence of external forcing, these interactions generate an antisymmetric stress component, ($\sigma_{xy}=-\sigma_{yx}$). At sufficiently large spinning rates, the resulting stresses fluidize an otherwise solid phase, much as an externally imposed shear melts a conventional amorphous solid beyond its yield stress.

Our results reveal a close connection between chirality-induced stresses and externally imposed shear. In the linear-response regime, transport coefficients are quantitatively described by a Green-Kubo relation once chirality is accounted for through an effective temperature. Beyond linear response, flow curves obtained for different shear and spinning rates collapse onto a single master curve when expressed in terms of the ratio, demonstrating that self-spinning acts as an intrinsic source of stress that competes directly with external deformation, thus inducing shear-thinning.   This unifies two seemingly distinct routes to fluidization within a common rheological framework.
Deep in the non-linear regime,  the interplay between shear and chirality generates phenomena that cannot be reduced to an effective temperature picture.
 When shear and spinning act with opposite handedness, stresses increase markedly and relax through the formation of layered, string-like structures that reorganize the flow. These states emerge from the competition between externally imposed deformation and chirality, revealing a regime where the handedness of the external/internal drive matters.

More broadly, our results place chiral fluids within the  framework of non-linear rheology of dense particle systems, where flow emerges from the competition between driving, dissipation and structural relaxation. 
We expect these findings to provide a useful reference for future studies of chiral active matter and odd mechanical metamaterials, where transverse interactions fundamentally reshape material properties and transport.

\textit{Acknowledgments-}
We thank for the access to the HPC resources of ReCas Data-Center  in Bari (Italy). L.M.C   and G.G
acknowledge support from INFN/FIELDTURB project and
Cineca-INFN agreement, providing access to HPC resources at CINECA.
D.L. acknowledges MCIU/AEI for financial support under
grant agreement PID2022-140407NB-C22. This research was supported in part by grant NSF PHY-2309135 to the Kavli Institute for Theoretical Physics (KITP). 
\bibliography{biblio.bib}

\section*{End Matter}
\setcounter{equation}{0}
\setcounter{figure}{0}
\renewcommand{\thefigure}{A\arabic{figure}}
\renewcommand{\theequation}{A\arabic{equation}}
\renewcommand{\thesection}{A}

\subsection{Stress self-correlation}
\begin{figure}[h]
    \centering
    \includegraphics[width=0.8\columnwidth]{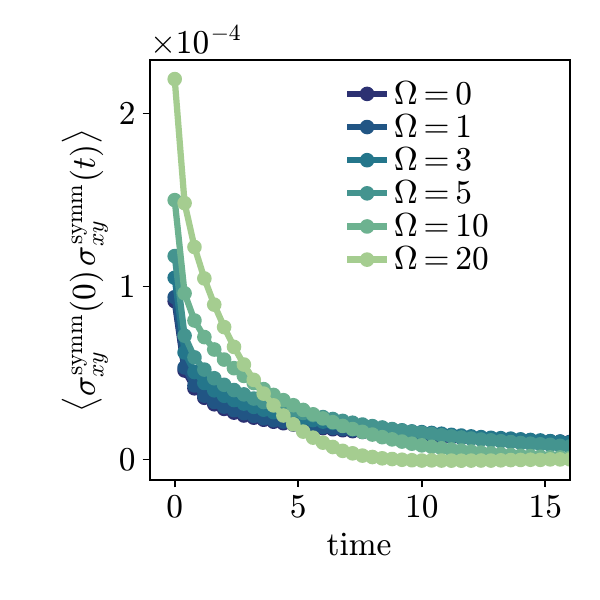}
    \caption{
    Autocorrelation of the symmetric off-diagonal stress $\langle \sigma_{xy}^{\rm s}(0)\,\sigma_{xy}^{\rm s}(t) \rangle$ measured in the unperturbed system for different values of activity.
    }
    \label{suppl2}
\end{figure}

In the main text, we compare the values of the viscosity extracted from the direct measurement of the Irving--Kirkwood stress tensor $\bm{\sigma}$ in the sheared system(Eq. 2 of the main text) with the ones computed from the generalized Green-Kubo formula, in the absence of shear: 
\begin{equation}
    \eta = \frac{A}{k_BT_{\rm eff}} \int_0^{+\infty} 
    \langle \sigma^{\rm s}(0)\,\sigma^{\rm s}(t) \rangle\, dt.
    \label{eq: green-kubo}
\end{equation}
This expression relies on the stress autocorrelation measured in the unperturbed steady state at each value of $\Omega$, and an independently determined effective temperature.
We measured the stress correlation for different values of chirality in a system of $N=65536$ particles and report the results in Fig.\ref{suppl2}.
Noticeably, the introduction of chirality leads to a growth in the equal-time autocorrelation, while appreciably shortening the relaxation time.
Interestingly, the integral resulting from the curves in Fig.\ref{suppl2}, together with the independently measured effective temperature, $k_B T_{eff}$, recovers the Newtonian plateau value of the viscosity measured in Fig.\ref{fig2}c of the main text.

\begin{figure*}[t!]
    \centering
    \includegraphics[width=1.0\linewidth]{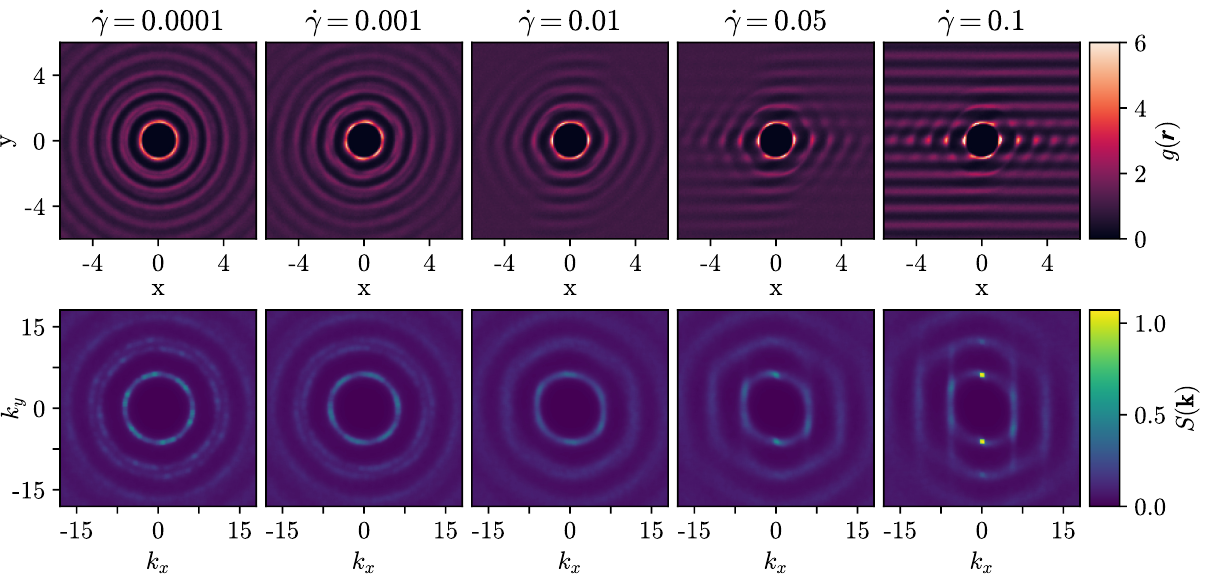}
    \caption{Pair correlation function (first row) and structure factor (second row) at fixed chirality, $\Omega=-5$, for different values of the shear rate $\dot\gamma$.}
    \label{em2}
\end{figure*}

\setcounter{equation}{0}
\setcounter{figure}{0}
\renewcommand{\thefigure}{B\arabic{figure}}
\renewcommand{\theequation}{B\arabic{equation}}
\renewcommand{\thesection}{B}
\subsection{Structure factor and layering criterion}

To characterize the crossover to the string-like regime, we use in the first row of Fig. \ref{em2} the full two-dimensional pair-correlation function. Numerically, we use a spatial binning $(\Delta x, \Delta y)$ and compute:
\begin{equation}
g(\bm{r})=\frac{A}{N(N-1) \Delta x \Delta y} \sum_{i\neq j} \int_x^{x+\Delta x}dx'\int_y^{y+\Delta y } dy' \;\delta(\bm{r}' - \bm{r}_{ij}) 
\end{equation}
with $\bm{r}=(x,y)$. 
We show in Fig. \ref{em2} the resulting $g(\bm{r})$ at five increasing values of the shear rate. At the smallest shear rates, the pair distribution function is essentially isotropic, with well defined concentric rings. As the shear rate increases, the rings become more distorted along the direction of the shear flow.
Finally, at the largest shear rate, the pair correlation function collapses onto stripes aligned with the flow, indicating the formation of well defined channels, giving rise to a string-like flow.
The crossover from the chiral liquid to the string-like flow is also captured in reciprocal space through the static structure factor defined as
\begin{equation}
    S(\bm{k}) = \frac{1}{N} \sum_{i,j} e^{i \bm{k}\cdot\bm{r_{ij}}}.
\end{equation}
The second row of Fig.~\ref{em2} displays the structure factor for the same fixed chiral activity, at increasing values of the shear rate. 
At low $\dot\gamma$ it retains the rotational symmetry of the liquid, with a circular ring at $|\bm{k}|\simeq 2\pi/\sigma_d$ set by the mean interparticle distance. As the shear rate grows, the rings turn increasingly anisotropic, displaying a growth along the velocity-gradient direction until at the largest $\dot\gamma$ it condenses into two bright peaks on the $k_y$ axis at $\bm{k}\simeq(0,\pm 2\pi/d_y)$, where $d_y$ is the characteristic distance set by the string-like patterns, structured into layers of particles.
The sharpening of this principal peak provides the quantitative criterion used to locate the crossover between the liquid to string-like flow regime indicated in the phase diagram of Fig.\ref{fig4}d: we mark the charge of regime by an order-of-magnitude increase in the height of the leading peak of $S(\bm{k})$, that coincides with the minimum of the bond-orientational order $|\psi_6|$. The consistency of these two criteria shows that the growth of $\overline{|\psi_6|}$ beyond $\dot\gamma^*$ reflects the onset of string-like flow rather than the restoration of hexatic order.

\end{document}